\documentclass[conference]{IEEEtran}
\pdfoutput=1
\newcommand{\mytitle}[0]{Graphulo Implementation of Server-Side \\ Sparse Matrix Multiply in the Accumulo Database}
\usepackage[  
  bookmarks=false,
  hyperfootnotes=false,
  hyperindex=false,
  hidelinks,  
  pdftitle={Graphulo Implementation of Server-Side Sparse Matrix Multiply in the Accumulo Database},  
  pdfauthor={Dylan Hutchison, Jeremy Kepner, Vijay Gadepally, Adam Fuchs}]{hyperref}
  
\usepackage{cite}
\usepackage{url}
\usepackage{tikz}
\usepackage{footnote}
\usepackage{balance}

\graphicspath{{.}{./img/}}
\usepackage{epstopdf}
\DeclareGraphicsExtensions{.eps,.pdf,.png}
\usepackage[cmex10]{amsmath}
\usepackage{todonotes}

\usepackage{dblfloatfix} 

\usepackage{threeparttable}
\usepackage{graphicx}
\usepackage{caption}
\usepackage{subcaption}
\usepackage[]{algorithm2e}
\usepackage{amsfonts} 
\usepackage{amssymb} 
\usepackage{authblk}

\newcommand{\matr}[1]{\mathbf{#1}} 
\newcommand{\tr}[0]{{\intercal}} 
\newcommand{\col}[0]{\colon\!}

\usepackage{mathtools}
\DeclarePairedDelimiter\ceil{\lceil}{\rceil}
\DeclarePairedDelimiter\floor{\lfloor}{\rfloor}

\usepackage{siunitx}
\usepackage{multirow}
\usepackage{adjustbox}
\usepackage{array}
\newcolumntype{R}[2]{%
    >{\adjustbox{angle=#1,lap=\width-(#2)}\bgroup}%
    l%
    <{\egroup}%
}

\makeatletter
\newcommand{\removelatexerror}{\let\@latex@error\@gobble}
\makeatother

\newlength{\algspace}
\newcommand{\matlab}{\textsc{Matlab}}



\begin{document}

\title{\mytitle{}}


\author[D. Hutchison et al.]
       {Dylan Hutchison,$^{\!{\dagger}{\S}*}\;$ Jeremy Kepner,$^{\!{\dagger}{\ddagger}{\diamond}*}\;$ Vijay Gadepally,$^{\!{\dagger}{\ddagger}*}\;$ Adam Fuchs$^+$ \\
         \\
         $^{\dagger}$MIT Lincoln Laboratory, 
         $^{\S}$University of Washington,\\
         $^{\ddagger}$MIT Computer Science \& AI Laboratory, 
         $^{\diamond}$MIT Mathematics Department, 
         $^+$Sqrrl, Inc. 
       }


%

\maketitle

{\let\thefootnote\relax\footnote{\hspace{-\parindent}Dylan Hutchison is the corresponding
    author, reachable at dhutchis@uw.edu.
}}
{\let\thefootnote\relax\footnote{*This material is based upon work
    supported by the National Science Foundation under Grant
    No. DMS-1312831. Opinions, findings, and conclusions or recommendations expressed in this material are those of the author(s) and do not necessarily reflect the views of the National Science Foundation.
}}
\setcounter{footnote}{0}
\begin{abstract}
The Apache Accumulo database excels at distributed storage
and indexing
and is ideally suited for storing graph data.
Many big data analytics
compute on graph data and persist their results back to the database.
These graph calculations are often best performed inside the database server. 
The GraphBLAS standard provides a compact and efficient basis for
a wide range of graph applications through a small number of sparse matrix operations. 
In this article, we discuss a server-side implementation of GraphBLAS sparse matrix multiplication
that leverages Accumulo's native, high-performance iterators.
We compare the mathematics and performance 
of inner and outer product implementations,
and show how an outer product implementation achieves optimal performance near
Accumulo's peak write rate.
We offer our work as a core component to the Graphulo library
that will deliver matrix math primitives for graph analytics within Accumulo.
\end{abstract}

\IEEEpeerreviewmaketitle

\let\stimes\times
\renewcommand{\times}[0]{{\,\stimes{}\,}}


\section{Introduction}
\label{sIntro}
%


The Apache Accumulo NoSQL database
was designed for high performance ingest and scans~\cite{sen2013benchmarking}. 
While fast ingest and scans solve some big data problems,
more complex scenarios involve performing tasks
such as data enrichment, graph algorithms and clustering analytics. These techniques
often require moving data from a database 
to a compute node. The ability to
compute directly in a database can lead to benefits including 
\emph{data locality}, \emph{infrastructure reuse} and \emph{selective access}. 

Accumulo administrators commonly create data locality 
by running server processes on the physical nodes where data is stored and cached.
Computing within Accumulo takes advantage of this locality 
by avoiding unnecessary network transfer,
effectively moving ``compute to data'' like a stored procedure,
in contrast to client-server models that move ``data to compute''.
Performing computation inside Accumulo also reuses its distributed infrastructure
such as write-ahead logging, fault-tolerant execution, and 
parallel load balancing of data.
In particular, Accumulo's infrastructure enables selective access to data along its indexed attributes (rows),
which enhances the performance of algorithms written with row access patterns.

There are a variety of ways to store graphs in Accumulo.  One common schema is
to store graphs as sparse matrices.  
Researchers in the GraphBLAS forum \cite{mattson2014standards} 
have identified a small set of kernels
that form a basis for matrix algorithms useful for graphs
when represented as sparse matrices.
This article presents Graphulo, an effort to realize the GraphBLAS primitives 
that enable algorithms using matrix mathematics directly in Accumulo servers \cite{gadepally2015gabb}.


In this paper we focus on Sparse Generalized Matrix Multiply (SpGEMM), the core kernel at the heart of GraphBLAS.
Many GraphBLAS primitives can be expressed in terms of
SpGEMM via user-defined multiplication and addition functions. 
SpGEMM can be used to implement a wide range of algorithms 
from graph search \cite{kepner2011graph} to table joins \cite{cohen2009mad} 
and many others (see introduction of \cite{bulucc2010highly}).

We call our implementation of SpGEMM in Accumulo \textsc{TableMult}, short for multiplication of Accumulo tables.
Accumulo tables have many similarities to sparse matrices, though a more precise mathematical definition is Associative Arrays 
\cite{kepner2014gabb}. For this work, we concentrate on
large distributed tables that may not fit in memory and use a streaming
approach that leverages Accumulo's built-in distributed infrastructure.

We are particularly interested in Graphulo for queued analytics~\cite{Reuther09cloudcomputing}, 
that is, analytics on selected table subsets.  
Queued analytics maximally leverage  databases
by quickly accessing subsets of interest, 
whereas whole-table analytics may perform better on parallel file systems such as Lustre or Hadoop.
We therefore prioritize smaller problems that require low latency
to enable analysts to explore graph data interactively.


We review Accumulo and its model for server-side computation, iterator stacks, 
in Section~\ref{sAccumuloIterators}.
We define matrix multiplication and compare inner and outer product methods
in Section~\ref{sMatMul}, settling on outer product for implementing TableMult.
We show TableMult's design as Accumulo iterators in Section~\ref{sTableMult}
and test TableMult's scalability with experiments in Section~\ref{sPerformance}.
We discuss related work, design alternatives and optimizations in Section~\ref{sDiscussion}
and conclude in Section~\ref{sConclusions}.



\subsection{Primer: Accumulo and its Iterator Stack}
\label{sAccumuloIterators}
Accumulo stores data in Hadoop RFiles as byte arrays indexed by key using (key, value) pairs called entries.
Keys decompose further into 5-tuples consisting of a row, column family, column qualifier, visibility and timestamp.
For simplicity, we focus on a 2-tuple key consisting of a row and column qualifier.
Entries belong to tables, which Accumulo divides into tablets and assigns to tablet servers.
Client applications write new entries via BatchWriters 
and retrieve entries sequentially via Scanners
or in parallel via BatchScanners.

Accumulo's server-side programming model runs an \emph{iterator stack} on tablets in range of a scan.
An iterator stack is a set of data streams originating
at Accumulo's data sources for a specific tablet (Hadoop RFiles and cached in-memory maps), 
converging together in merge-sorts,
flowing through each iterator in the stack and at the end, sending entries to the client.
Iterators themselves are Java classes implementing the SortedKeyValueIterator (SKVI) interface.

Developers add custom logic for server-side computation
by writing new iterators and plugging them into the iterator stack.
In return for fitting their computation in the SKVI paradigm, developers gain
distributed parallelism for free as Accumulo runs their iterators on relevant tablets simultaneously.


SKVIs are reminiscent of built-in Java iterators in that they hold state 
and emit one entry at a time until finished iterating.
However, they are more powerful than Java iterators in that they can seek to arbitrary positions
in the data stream. 
They also have two constraints: 
the end of the iterator stack should emit entries in sorted order,
and iterators must not maintain volatile state such as threads, open files or sockets
because Accumulo may destroy, re-create and re-seek an iterator stack
between function calls without allowing time to clean up.



Iterators are most commonly used for ``reduction'' operations that transform
or eliminate entries passing through.  The Accumulo community generally discourages ``generator'' iterators 
that emit new entries not present in data sources 
because they are easy to misuse and violate SKVI constraints by emitting entries out of order or 
relying on volatile state.
In this work, we suggest a new pattern for iterator usage as a conduit for client write operations 
that achieves the benefits of generator iterators while avoiding their constraints.


\section{TableMult Design}
\label{sDesign}

\subsection{Matrix Multiplication}
\label{sMatMul}
Given matrices $\matr{A}$ of size $N \times M$, $\matr{B}$ of size $M \times L$,
and operations $\oplus$ and $\otimes$ for element-wise addition and multiplication,
the matrix product $\matr{C} = \matr{A} \,{\oplus}.{\otimes}\, \matr{B} $, or more shortly $\matr{C} = \matr{A}\matr{B}$,
defines entries of result matrix $\matr{C}$ as 
\[ \matr{C}(i,j) = \bigoplus_{k=1}^M \matr{A}(i,k) \otimes \matr{B}(k,j) \]
We call intermediary results of $\otimes$ operations \emph{partial products}.

For the sake of sparse matrices, we only perform $\oplus$ and $\otimes$ when both operands are nonzero,
an optimization stemming from requiring that 0 is an additive identity such that $a \oplus 0 = 0 \oplus a = a$,
and that 0 is a multiplicative annihilator such that $a \otimes 0 = 0 \otimes a = 0$.
Without these conditions, zero operands could generate nonzero results that destroy sparsity.


We study two well known patterns for computing matrix multiplication:
inner product and outer product \cite{kruskal1989techniques}. They differ in the order in which they perform
the $\otimes$ and $\oplus$ operations.  The more common inner product approach runs the following: 

\removelatexerror
\begin{algorithm}[H]
\vspace{\algspace}
\SetKwBlock{fore}{for}{} 
\SetKw{emit}{emit}
\fore($i = 1\col N$){
\fore($j = 1\col L$){
{$\matr{C}(i,j) \mathrel{\oplus}= \matr{A}(i,:)  \matr{B}(:,j)$}
}}
\vspace{\algspace}
\end{algorithm}

\noindent 
performing inner product on vectors.
For easier comparison, we rewrite the above approach with summation deferred as:

\removelatexerror
\begin{algorithm}[H]
\vspace{\algspace}
\SetKwBlock{fore}{for}{} 
\SetKw{emit}{emit}
\fore($i = 1\col N$){
\fore($j = 1\col L$){
\fore($k = 1\col M$){
{$\matr{C}(i,j) \mathrel{\oplus}= \matr{A}(i,k) \otimes \matr{B}(k,j)$}
}}}
\vspace{\algspace}
\end{algorithm}

Inner product has the advantage of generating entries in sorted order:
the third-level loop generates all partial products needed 
to compute a particular element $\matr{C}(i,j)$ consecutively.
The $\oplus$ applies immediately after each third-level loop to obtain an element in $\matr{C}$.
Inner product is therefore easy to ``pre-sum,'' an Accumulo term for applying a Combiner
locally before sending entries to a remote but globally-aware table Combiner.
Emitting sorted entries also facilitates inner product use in standard iterator stacks and easier operation pipelining.

Despite inner product's order-preserving advantages, 
outer product performs better for sparse matrices 
because it passes through $\matr{A}$ and $\matr{B}$ only once 
\cite{burkhardt2013big}\cite{burkhardt2014asking}.  
Inner product's second-level loop repeats
a scan over all of $\matr{B}$ for each row of $\matr{A}$.
Under our assumption that we cannot fit $\matr{B}$ entirely in memory,
multiple passes over $\matr{B}$ translate to multiple Accumulo scans that each require a disk read.
We found in performance tests that an outer product approach performs an order of magnitude better than an inner product approach.

The outer product approach runs the following:

\removelatexerror
\begin{algorithm}[H]
\vspace{\algspace}
\SetKwBlock{fore}{for}{} 
\SetKw{emit}{emit}
\fore($k = 1\col M$){
{$\matr{C} \mathrel{\oplus}= \matr{A}(:,k) \matr{B}(k,:)$}
}
\vspace{\algspace}
\end{algorithm}

\noindent 
performing outer 
product on vectors that corresponds to many elements of $\matr{C}$. 
Unfolding outer product reveals them 
as:

\removelatexerror
\begin{algorithm}[H]
\vspace{\algspace}
\SetKwBlock{fore}{for}{} 
\SetKw{emit}{emit}
\fore($k = 1\col M$){
\fore($i = 1\col N$){
\fore($j = 1\col L$){
{$\matr{C}(i,j) \mathrel{\oplus}=  \matr{A}(i,k) \otimes \matr{B}(k,j)$}
}}}
\vspace{\algspace}
\end{algorithm}

Compared to inner product, outer product moves the $k$ loop
above the $i$ and $j$ loops that determine position in $\matr{C}$.
The switch results in generating partial products out of order.


On the other hand, outer product only requires a single pass over both input matrices.
This is because the top-level $k$ loop fixes a dimension of both $\matr{A}$ and $\matr{B}$.
Once we finish processing a column of $\matr{A}$ and row of $\matr{B}$,
we never need read them again.

In terms of memory usage, outer product works best when either the matching row or column fits in memory.
If neither fits, we could run the algorithm 
with a ``no memory assumption'' streaming approach
by re-reading $\matr{B}$'s rows while streaming through $\matr{A}$'s columns 
(or vice versa by symmetry of $i$ and $j$),
perhaps at the cost of extra disk reads.

Because $k$ runs along $\matr{A}$'s second dimension 
and Accumulo uses row-oriented data layouts, we implement 
TableMult to operate on $\matr{A}$'s transpose $\matr{A^\tr}$.

\subsection{TableMult Iterators}
\label{sTableMult}

TableMult uses three iterators placed on a BatchScan of table $\matr{B}$:
RemoteSourceIterator, TwoTableIterator and RemoteWriteIterator.
A BatchScanner directs Accumulo to run the iterators on $\matr{B}$'s tablets in parallel.

The key idea behind the TableMult iterators is that they divert normal dataflow by opening a BatchWriter,
redirecting entries out-of-band to $\matr{C}$ via
Accumulo's unsorted ingest channel. 
The scan itself emits no entries except for a small number of ``monitoring entries'' 
that inform the client about TableMult progress.
We permit multi-table iterator dataflow by opening Scanners 
that read remote Accumulo tables out-of-band.
Scanners and BatchWriters are standard tools for Accumulo clients; 
by creating them inside iterators, we enable client-side processing patterns
within tablet servers.

Underlying our use of iterators, Scanners and BatchWriters are Accumulo's 
standing thread pools. 
Thread pools fulfill our low latency requirement by 
executing upon receiving a request 
at no more expense than a context switch.
Scaling up may require tuning thread pool size
to balance thread contention.

\begin{figure}[htb]
\centering
\includegraphics[width=3.0in]{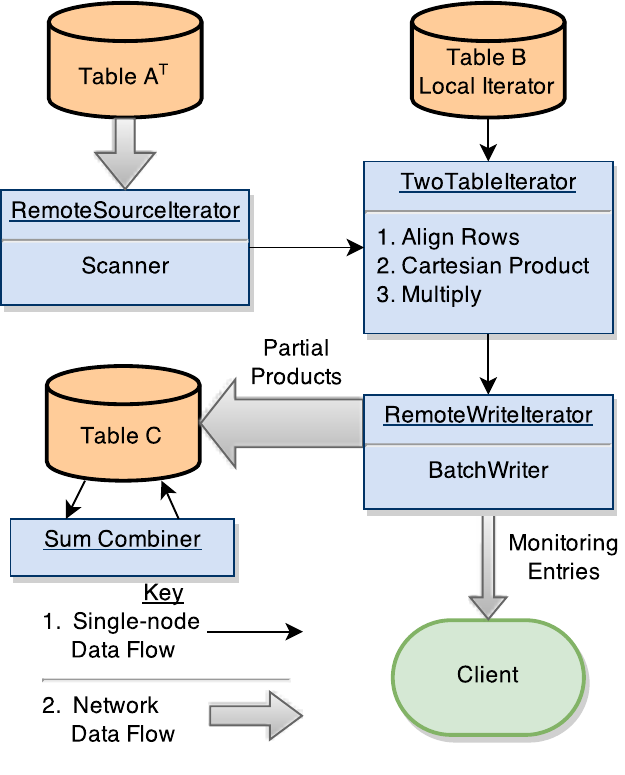}
\caption{Data flow through the TableMult iterator stack}
\label{fIteratorStackSpGEMM}
\end{figure}

We illustrate TableMult's data flow in Figure~\ref{fIteratorStackSpGEMM},
placing a Scanner on table $\matr{A^\tr}$
and a BatchWriter on result table $\matr{C}$.

\subsubsection{RemoteSourceIterator}
RemoteSourceIterator scans an Accumulo table
(not necessarily in the same cluster) 
using credentials passed from the client through iterator options.

We also use iterator options to specify row and column subsets, 
encoding them in a string format similar to that in D4M~\cite{kepner2012dynamic}.
Row subsets are straightforward since Accumulo uses row-oriented indexing.
Column subsets can be implemented with filter iterators
but do not improve performance since Accumulo must read every column from disk.
We encourage users to maintain a transpose table
using strategies similar to the D4M Schema~\cite{kepner2013d4m}
for cases requiring column indexing.

Multiplying table subsets is crucial for queued analytics on selected rows.
However for simpler performance evaluation, 
our experiments in Section~\ref{sPerformance} multiply whole tables.

\subsubsection{TwoTableIterator}
TwoTableIterator reads from two iterator sources, one for $\matr{A^\tr}$ and one for $\matr{B}$,
and performs the core operations of the outer product algorithm in three phases:
\begin{enumerate}
\item Align Rows.  Read entries from $\matr{A^\tr}$ and $\matr{B}$ until they advance to a matching row
or one runs out of entries. We skip non-matching rows 
since they would multiply with an all-zero row that, by Section~\ref{sMatMul}'s assumptions,
generate all zero partial products.
\item Cartesian product. Read both matching rows into an in-memory data structure. 
Initialize an iterator that emits pairs of entries from the rows' Cartesian product.
\item Multiply. Pass pairs of entries to $\otimes$ and emit results. 
\end{enumerate}

A client defines $\otimes$ by specifying a class 
that implements a multiply interface.
For our experiments we implement $\otimes$ as java.math.BigDecimal multiplication,
which guarantees correctness under large or precise real numbers.
BigDecimal decoding did not noticeably impact performance.

\subsubsection{RemoteWriteIterator}
RemoteWriteIterator writes entries to a remote Accumulo table using a BatchWriter. 
Entries do not have to be in sorted order because Accumulo sorts incoming entries as part of its
 ingest process. 

Barring extreme events such as exceptions in the iterator stack or thread death,
we designed RemoteWriteIterator to maintain correctness, such that entries generated from
its source write to the remote table once.
We accomplish this by performing all BatchWriter operations within a single function call
before ceding thread control back to the tablet server.  

A performance concern remains when multiplying a subset of the input tables' rows 
that consists of many disjoint ranges, such as one million ``singleton'' ranges spanning one row each.
It is inefficient to flush the BatchWriter before returning from each seek call, which happens once per 
disjoint scan range. 
We accommodate this case by ``transferring seek control'' 
from the tablet server to RemoteWriteIterator 
via the same strategy used in RemoteSourceIterator for seeking within an iterator.

We include an option to BatchWrite $\matr{C}$'s transpose $\matr{C^\tr}$ in place of or alongside $\matr{C}$. 
Writing $\matr{C^\tr}$ facilitates chaining TableMults together
and maintenance of transpose indexing.

\subsubsection{Lazy $\oplus$}
We lazily sum partial products by placing a Combiner subclass implementing BigDecimal addition 
on table $\matr{C}$ at scan, minor and major compaction scopes.
Thus, $\oplus$ occurs sometime after RemoteWriteIterator writes partial products to $\matr{C}$
yet necessarily before entries from $\matr{C}$ may be seen such that we always achieve correctness.
Summation could happen when Accumulo flushes $\matr{C}$'s entries cached in memory to a new RFile, 
when Accumulo compacts RFiles together, or when a client scans $\matr{C}$. 

The key algebraic requirement for implementing $\oplus$ inside a Combiner
is that $\oplus$ must be associative and commutative.
These properties allow us to apply $\oplus$ to subsets of a result element's partial products 
and to any ordering of them, which is chaotic by outer product's nature.
If we truly had an $\oplus$ operation that required seeing all partial products at once,
we would have to either gather partial products at the client or initiate a full major compaction.

\subsubsection{Monitoring}
RemoteWriteIterator never emits entries to the client by default. 
One downside of this approach is that clients cannot precisely track progress of a TableMult operation,
which may frustrate users expecting a more interactive computing experience.
Clients could query the Accumulo monitor for read/write rates 
or prematurely scan partial products written to $\matr{C}$, but both approaches are too coarse.

We therefore implement a monitoring option that emits a value
containing the number of entries TwoTableIterator processed
at a client-set frequency.
RemoteWriteIterator emits monitoring entries at ``safe'' points, that is,
points at which we can recover the iterator stack's state 
if Accumulo destroys, re-creates and re-seeks it.
Stopping after emitting the last value in the outer product of two rows is safe 
because we place the last value's row key in the monitoring key and know, 
in the event of an iterator stack rebuild, to proceed to the next matching row.
We may succeed in stopping during an outer product 
by encoding more information in the monitoring key.



 




\section{Performance}
\label{sPerformance}

We evaluate TableMult with two variants of an experiment. 
First we measure the rate of computation as problem size increases.
We define problem size as number of rows in random input graphs 
represented as adjacency tables
and rate of computation as number of partial products processed per second.
Second we repeat the experiment for a fixed size problem with all tables split into two tablets,
allowing Accumulo to scan and write to them in parallel.



We compare Graphulo TableMult performance to D4M~\cite{kepner2012dynamic} as a baseline because 
a user with one client machine's best alternative is reading input graphs from Accumulo, 
multiplying them at the client, and inserting the result back into Accumulo.

D4M stores tables as Associative Array objects in \matlab{}.  
Because Assoc Array multiplication runs fast
by calling \matlab{}'s in-memory sparse matrix functions, 
D4M bottlenecks on reading data from Accumulo and especially on writing back results,
despite its capacity for high speed Accumulo reads and writes~\cite{kepner2014achieving}.
We consequently expect TableMult to outperform D4M 
because TableMult avoids transferring data out of Accumulo for processing. 

We also expect TableMult to succeed on larger graph sizes than D4M because TableMult
uses a streaming outer product algorithm that does not store input tables in memory.
An alternative D4M implementation would mirror TableMult's streaming outer product algorithm,
enabling D4M to run on larger problem sizes at potentially worse performance.
We therefore imagine the whole-table D4M algorithm as an upper bound on the best performance 
achievable when multiplying Accumulo tables outside Accumulo's infrastructure.

We use the Graph500 unpermuted power law graph generator \cite{bader2006designing} 
to create random input tables,
such that both tables' first row have high degree (number of columns) 
and subsequent rows exponentially decrease in degree.
The common power law structure correlates the input tables, 
which leads to denser result tables than if we were to permute the input tables
but does not otherwise affect multiplication behavior.
The generator takes SCALE and EdgesPerVertex parameters, creating graphs with 2\textsuperscript{SCALE} 
rows and EdgesPerVertex $\times$ 2\textsuperscript{SCALE} entries.
We fix EdgesPerVertex to 16 and use SCALE to vary problem size. 

The following procedure outlines our performance experiment for a given SCALE and either one or two tablets.
\begin{enumerate}
\item Generate two graphs with different random seeds and insert them into Accumulo as adjacency tables via D4M.
\item In the case of two tablets, identify an optimal split point for each input graph
and set the input graphs' table splits equal to that point.
``Optimal'' here means a split point that evenly divides an input graph into two tablets.
\item \label{ePreSplit1} Create an empty output table. For two tablets, pre-split it with 
an optimal input split position recorded from a previous multiplication run.
\item \label{ePreSplit1Compact} Compact the input and output tables 
so that Accumulo redistributes the tables' entries into the assigned tablets.
\item Run and time Graphulo TableMult multiplying the transpose of the first input table with the second.
\item Create, pre-split and compact a new result table for the D4M comparison 
as in step~\ref{ePreSplit1} and~\ref{ePreSplit1Compact}.
\item Run and time the D4M equivalent of TableMult:
 \begin{enumerate}
 \item Scan both input tables into D4M Associative Array objects in \matlab{} memory.
 \item Convert the string values from Accumulo into numeric values for each Associative Array.
 \item Multiply the transpose of the first Associative Array with the second.
 \item Convert the result Associative Array back to String values and insert them into Accumulo.
 \end{enumerate}
\end{enumerate}

We conducted the experiments on a Ubuntu Linux laptop with 16GB RAM and two dual-core Intel i7 processors
. Using single-instance Accumulo 1.6.1, Hadoop 2.6.0 and ZooKeeper 3.4.6,
we allocated 2GB of memory to an Accumulo tablet server initially
(allowing growth in 500MB steps),
1GB for native in-memory maps and 256MB for data and index cache.

We chose not to use more than two tablets per table because more threads would run
than the laptop could handle.  Each additional tablet can potentially add the following threads:
\begin{enumerate}
\item Table $\matr{A}^\tr$ server-side scan thread;
\item Table $\matr{A}^\tr$ client-side scan thread,

$\quad$ running from RemoteSourceIterator;
\item Table $\matr{B}$ server-side scan/multiply thread,

$\quad$ running a TableMult iterator stack;
\item Table $\matr{B}$ client-side scan thread, 

$\quad$ running from the initiating client, mostly idle;
\item Table $\matr{C}$ server-side write thread;
\item Table $\matr{C}$ client-side write thread,

$\quad$ running from RemoteWriteIterator; and
\item Table $\matr{C}$ server-side minor compaction threads,

$\quad$ running with a Combiner implementing $\oplus$.
\end{enumerate}

\begin{figure}[t]
\centering
\includegraphics[width=\linewidth]{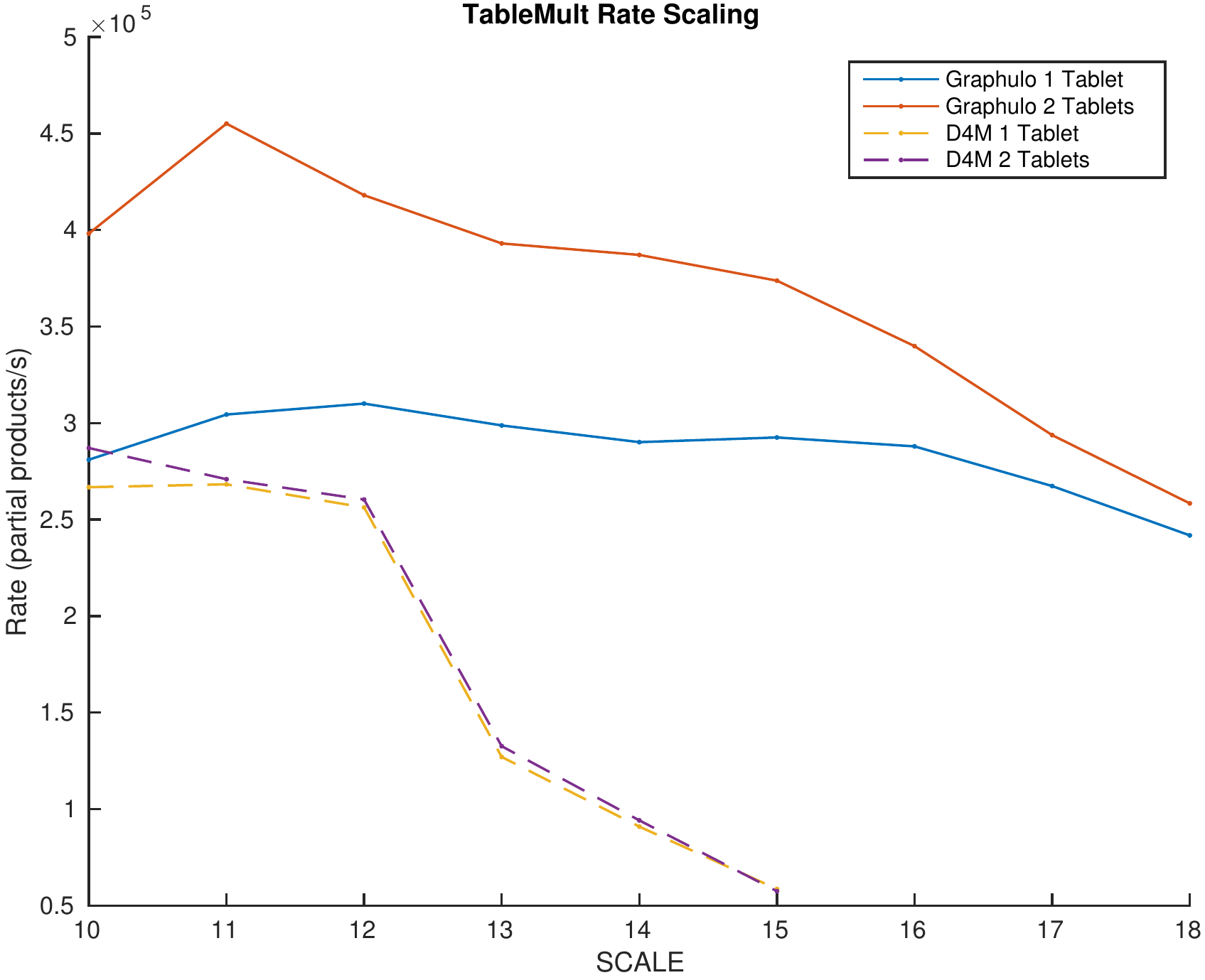}
\caption{TableMult Processing Rate vs. Input Table Size}
\label{fTableMultPerf}
\end{figure}

We show table $\matr{C}$ sizes and experiment timings in Table~\ref{tResultsParams}
and plot them in Figure~\ref{fTableMultPerf}.
We could not run the D4M comparison past SCALE 15 because $\matr{C}$ does not fit in memory.

For the scaled problem, the best results we could achieve are flat horizontal lines, 
indicating that we maintain the same level of operations per second as problem size increases.

One reason we see a downward rate trend at larger problem sizes is that Accumulo
needs to minor compact table $\matr{C}$ in the middle of a TableMult. 
The compactions trigger flushes to disk along with 
the $\oplus$ Combiner that sums partial products written to $\matr{C}$ so far, 
neither of which we include 
in rate measurements. 

For the fixed size problem, the best results we could achieve are two-tablet rates at
double the one-tablet rates at every problem size.
Our experiment shows that Graphulo two-tablet rates perform up to 1.5x better
than one-tablet rates at lower SCALEs. 
We attribute TableMult's shortfall to high processor contention for the laptop's four cores as a result of 
the 14 threads that may run concurrently when each table has two tablets; in fact,
processor usage hovered near 100\% for all four cores throughout the two-tablet experiments.
We expect better scaling when running our experiment 
in larger Accumulo clusters that can handle more degrees of parallelism.


\begin{table*}[tb]
\centering                                                                                                           
\caption{Output Table $\matr{C}$ Sizes and Experiment Timings}
\label{tResultsParams}
\begin{threeparttable}[c]
\addtolength{\tabcolsep}{-0.5pt}  
\begin{tabular}{r|ll|ll|ll|ll|ll}
\multirow{2}{1.75em}{\adjustbox{angle=30,lap=\width-3.75em}{SCALE}} & \multicolumn{2}{c|}{Entries in Table $\matr{C}$} & \multicolumn{2}{c|}{Graphulo 1 Tablet} & \multicolumn{2}{c|}{D4M 1 Tablet} & \multicolumn{2}{c|}{Graphulo 2 Tablets} & \multicolumn{2}{c}{D4M 2 Tablets} \\
 & PartialProducts\hspace{-0.75em} & AfterSum & Time (s) & Rate (pp/s) & Time (s) & Rate (pp/s) & Time (s) & Rate (pp/s) & Time (s) & Rate (pp/s) \\             
\hline
10 & \num{804989.000} & \num{269404.000} & \num{2.865} & \num{281012.707} & \num{3.018} & \num{266771.720} & \num{2.022} & \num{398174.309} & \num{2.804} & \num{287060.355} \\             
\hline                                                                                                                                                                                                          
11 & \num{2361580.000} & \num{814644.000} & \num{7.758} & \num{304413.622} & \num{8.803} & \num{268259.547} & \num{5.189} & \num{455121.509} & \num{8.718} & \num{270898.575} \\            
\hline                                                                                                                                                                                                          
12 & \num{6816962.000} & \num{2430381.000} & \num{21.984} & \num{310090.248} & \num{26.601} & \num{256270.986} & \num{16.307} & \num{418039.002} & \num{26.182} & \num{260366.279} \\       
\hline                                                                                                                                                                                                          
13 & \num{19111689.000} & \num{7037007.000} & \num{63.969} & \num{298766.256} & \num{150.475} & \num{127009.402} & \num{48.623} & \num{393059.423} & \num{144.156} & \num{132575.978} \\    
\hline                                                                                                                                                                                                          
14 & \num{52656204.000} & \num{20029427.000} & \num{181.506} & \num{290106.916} & \num{579.243} & \num{90905.237} & \num{136.025} & \num{387107.096} & \num{559.271} & \num{94151.551} \\   
\hline                                                                                                                                                                                                          
15 & \num{147104084.000} & \num{58288789.000} & \num{502.864} & \num{292532.774} & \num{2510.389} & \num{58598.135} & \num{393.573} & \num{373765.880} & \num{2559.243} & \num{57479.523} \\
\hline                                                                                                                                                                                                          
16 & \num{400380031.000} & \num{163481262.000} & \num{1390.612} & \num{287916.484} &  &  & \num{1178.111} & \num{339849.273} &  &  \\                                                               
\hline                                                                                                                                                                                                          
17 & \num{1086789275.000} & \num{459198683.000} & \num{4064.990} & \num{267353.526} &  &  & \num{3699.671} & \num{293752.983} &  &  \\                                                              
\hline                                                                                                                                                                                                          
18 & $2.94 \times 10^9$
& \num{1280878452.000} & \num{12148.744} & \num{241798.621} &  &  & \num{11369.009} & \num{258382.204} &  &  \\
\end{tabular}
\end{threeparttable}
\end{table*}

\renewcommand{\times}[0]{\stimes}


\section{Discussion}
\label{sDiscussion}

\subsection{Related Work} 
Bulu\c{c} and Gilbert studied message passing algorithms for SpGEMM
such as Sparse SUMMA, most of which use 2D block decompositions \cite{buluc2012parallel}.
Unfortunately, 2D decompositions are difficult in Accumulo 
and message passing even more so.
In this work, we use Accumulo's native 1D decomposition along rows 
and do not rely on tablet server communication
apart from shuffling partial products of $\matr{C}$ via BatchWriters.

Our outer product method could have been implemented on Hadoop MapReduce 
 or its successor YARN \cite{vavilapalli2013apache}.
There is a natural analogy from TableMult to MapReduce:
the map phase scans rows from $\matr{A^\tr}$ and $\matr{B}$
and generates a list of partial products from TwoTableIterator;
the shuffle phase sends partial products to correct tablets of $\matr{C}$ via BatchWriters;
and the reduce phase sums partial products using Combiners.
Examining the conditions on which MapReduce 
reading from and writing to Accumulo's RFiles directly
can outperform Accumulo-only solutions
is worthy future work.


A common Accumulo pattern is to 
scan and write from multiple clients in parallel;
in fact, researchers obtained 
considerably high insert rates using parallel client strategies \cite{kepner2014achieving}.
We chose to build Graphulo as a service within Accumulo
instead of assuming a multiple client capability,
such that Graphulo is as accessible as possible to diverse client environments.

The strategy in \cite{kepner2014achieving} 
also used 
tablet location information
to determine where clients could write locally.
Knowing tablet-to-tablet-server assignment could likewise aid Graphulo, 
not only to minimize network traffic
but also
 to partly eliminate Apache Thrift RPC serialization,
which prior work has shown is a bottleneck for scans 
when iterator processing is light \cite{sawyer2013understanding}.
Such an enhancement would access a local tablet server by method call 
in place of Scanners and BatchWriters.


The Knowledge Discovery Toolkit (KDT) distributed-memory Python graph library
offers sparse matrix multiplication in a similar design as Graphulo's~\cite{bulucc2013high}.
Both support custom addition, multiplication and filter operators written 
in a high level language.
They differ in that Graphulo targets the Accumulo infrastructure
which is IO-bound, 
in contrast to the KDT which is compute-bound. 
Graphulo therefore gains less from code generation techniques on its Java iterator kernels, 
whereas the KDT uses the SEJITS technique~\cite{catanzaro2009sejits} to translate 
Python kernels into C++ 
for callback by KDT's underlying Combinatorial BLAS library~\cite{bulucc2011combinatorial},
thereby raising performance from compute- to memory bandwidth-bound 
at the expense of restricting operator expressiveness to a DSL.

\subsection{Design Alternative: Inner-Outer Product Hybrid}

\begin{figure}[b]
\centering
\includegraphics[width=\linewidth]{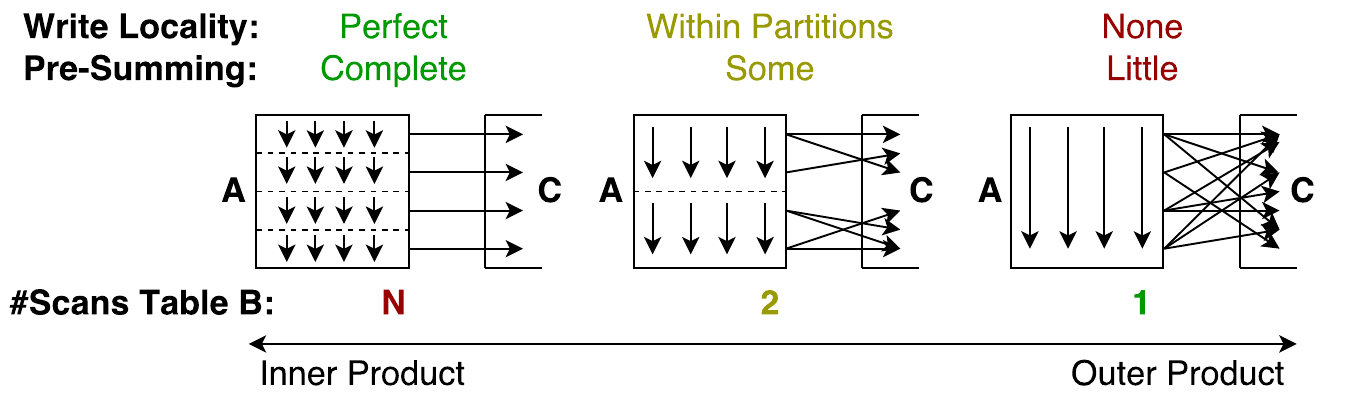}
\caption{Tradeoffs between Inner and Outer Product}
\label{fInnerOuterSpectrum}
\end{figure}

It is worth reconsidering the inner product method from our initial design
because it has an opposite performance profile as 
Figure~\ref{fInnerOuterSpectrum}'s left and right depict: 
inner product bottlenecks on scanning whereas outer product bottlenecks on writing.
At the expense of multiple passes over input matrices, inner product emits 
partial products in order and immediately pre-summable,
reducing the number of entries written to Accumulo to the minimum possible.
Outer product reads inputs in a single pass
but emits entries out of order and has little chance to pre-sum, 
instead writing individual partial products to $\matr{C}$.
Table~\ref{tResultsParams} quantifies that outer product writes
2.5 to 3 times more entries than inner product for power law inputs.
In the worst case, multiplying a fully dense $N \times M$ with an $M \times L$ matrix,
outer product emits $M$ times more entries than inner product.

Is it possible to blend inner and outer product SpGEMM methods,
choosing a middle point in Figure~\ref{fInnerOuterSpectrum}
with equal read and write bottlenecks for overall greater performance?
In 
the following generalization, 
partition parameter $P$ varies behavior between
inner product at $P=N$ and outer product at $P=1$:

\removelatexerror
\begin{algorithm}[H]
\vspace{\algspace}
\SetKwBlock{fore}{for}{} 
\SetKw{emit}{emit}
\fore($p = 1\col P$){
\fore($k = 1\col M$){
\fore({$i = \left( \floor*{\dfrac{(p-1)N}{P}}+1 \right) \col \floor*{\dfrac{p N}{P}}$}){
\fore($j = 1\col L$){
{$\matr{C}(i,j) \mathrel{\oplus}= \matr{A}(i,k) \otimes \matr{B}(k,j)$}
}}}}
\vspace{\algspace}
\end{algorithm}

The hybrid algorithm runs $P$ passes through $\matr{B}$,
each of which has write locality to a vertical partition of $\matr{C}$ of size $N/P \times L$.
Pre-summing ability likewise varies inversely with $P$, 
though actual pre-summing depends on
$\matr{A}$ and $\matr{B}$'s  sparsity distribution
as well as how many positions of $\matr{C}$ the TableMult iterators cache.
Figure~\ref{fInnerOuterSpectrum}'s center depicts the $P=2$ case.

A challenge for any hybrid algorithm is mapping it to Accumulo infrastructure.
We chose outer product because it more naturally fits Accumulo, 
using iterators for one-pass streaming computation, 
BatchWriters to handle unsorted entry emission and Combiners to defer summation.
The above hybrid algorithm resembles 2D block decompositions,
and so maximizing its performance may be challenging 
given limited data layout control and unknown data distribution.

Nevertheless, possible design criteria are to select a small $P$ to minimize passes through $\matr{B}$,
while also choosing $P$ large enough so that $\ceil{NL/P}$ entries fit in memory
(dense matrix worst case), which guarantees complete pre-summing.
The latter criterion may be relaxed with decreasing matrix density.

\subsection{TableMult in Algorithms}

Several optimization opportunities exist for TableMult as a primitive in larger algorithms.
Given row $\matr{A}$ of starting vertices and graph adjacency matrix $\matr{B}$, 
suppose we wish to union the vertices reached in two steps from those in $\matr{A}$ 
into $\matr{A}$ via the program
 $\matr{C} = \matr{AB}; \matr{D} = \matr{CB}; \matr{A} \mathrel{\oplus}= \matr{D}$,
as one way of calculating $\matr{A} \mathrel{\oplus}= \matr{AB}^2$ via TableMult calls.
Such calculations are useful for finding vertices reachable in an even number of steps.
We would save two round trips to disk if we could mark $\matr{C}$ and $\matr{D}$ as 
``temporary tables,'' i.e. tables intermediate to an algorithm that should be held in memory 
and not written to Hadoop if possible.
Combiners in TableMult do enable one optimization:
summing $\matr{CB}$ into $\matr{A}$ directly by rewriting the program as 
$\matr{C} = \matr{AB}; \matr{A} \mathrel{\oplus}= \matr{CB}$.

A \emph{pipelining} optimization streams entries from a TableMult 
to computations taking its result as input. 
Outer product pipelining is difficult
because it cannot guarantee writing every partial product for a particular element 
 to $\matr{C}$ until it finishes,
whereas inner product's complete pre-summing emits elements ready for use downstream.
More ambitiously, \emph{loop fusion} merges iterator stacks 
for successive computations into one. 

Optimizing computation on NoSQL databases is challenging in the general case because
NoSQL databases typically avoid query planner features 
customary of SQL databases in exchange for raw performance.
NewSQL databases aim in part to achieve the best of both worlds---performance and query planning \cite{grolinger2013data}.
We aspire to make a small step for Accumulo in the direction of NewSQL with current Graphulo research.

\section{Conclusions}
\label{sConclusions}

In this work we showcase the design of TableMult, a Graphulo server-side implementation of the 
SpGEMM GraphBLAS matrix math kernel in the Accumulo database.
We compare inner and outer product approaches and show how outer product 
better fits Accumulo's iterator model.  The implementation shows excellent single node performance, 
achieving write rates near 400,000 per second, 
which is consistent with the single node peak write rate for Accumulo \cite{kepner2014achieving}.
Performance experiments show good scaling for scaled problem sizes and suggest good scaling for fixed size problems,
but these require additional experiments on a larger cluster to confirm.

In addition to topics from Section~\ref{sDiscussion}'s discussion, 
future research efforts include
implementing the remaining GraphBLAS kernels, 
developing graph algorithms that use the Graphulo library
and delivering to the Accumulo community.

\bibliographystyle{IEEEtran}

\bibliography{10_bibliography}

\balance

\end{document}